\documentclass[amsmath,amssymb,twocolumn]{revtex4}
\usepackage{dcolumn}
\usepackage{bm}
\usepackage{amsmath,mathrsfs}
\usepackage{graphicx,epsfig}
\usepackage{epsf}
\usepackage{color}

\begin{document}


\title{Asymptotically flat  wormhole solutions with variable equation-of-state parameter}

\author{F. Parsaei}\email{fparsaei@gmail.com}
\author{S. Rastgoo}\email{rastgoo@sirjantech.ac.ir}
\affiliation{Physics Department , Sirjan University of Technology, Sirjan 78137, Iran}

\date{\today}


\begin{abstract}

In this paper, we study exact wormhole solutions in the framework of general relativity with a general equation of state that reduced to a linear equation of state asymptotically. By considering a special shape function, we find  classes of solutions which are asymptotically flat.  We study the violation of NEC as the main ingredient in the wormhole physics. We investigate the possibility of finding wormhole solutions with asymptotically different state parameter. We show that in principle, wormhole with a vanishing redshift function and the selected shape function, cannot satisfy NEC at large distance from wormhole. We present solutions which have the positive total amount of mater in the "volume integral quantifier" method. For this class of solutions, fluid near the wormhole throat is in the phantom regime and at some $r=r_{2}$, the phantom regime is connected to a dark energy regime. Thus, we need small amount of exotic matter to construct wormhole solutions.


\end{abstract}

\maketitle

\section{Introduction}

Historically, Flamm has suggested the concept of wormhole \cite{flamm}. After Flamm, a similar construction, famous as Einstein-Rosen bridge, was presented by Einstein and Rosen \cite{Rosen}. Finally, Morris and Thorne have introduced  the traversable wormhole for interstellar or time travel\cite{WH}. In other words, wormhole is an exact solution of Einstein field equations which can connect two universes or two distant parts of the same universe. The main ingredient in the wormhole theory is the violation of classical energy conditions \cite{Visser}.  The matter which violates null energy condition (NEC) specified by $T_{\mu\nu}k^{\mu}k^{\nu}\geq0$, in which $k^{\mu}$ is any null vector and $T_{\mu\nu}$ stress-energy tensor,  is called exotic. Someone needs exotic matter to construct wormhole \cite{Visser}. Since ordinary and laboratory matters obey energy conditions, many authors  try to solve the problem of exotic matter in studying wormhole theory. In scalar-tensor theories, scalar fields play the role of exotic matter, in particular, wormhole solutions can be obtained if  scalar fields are phantom \cite{Br}. Many classes of attempts have been based on the modified gravity theories such as $f(R)$ gravity\cite{tahereh}, curvature matter coupling \cite{modgravity2b} and brane-world\cite{brane1,brane}. In these theories an effective stress-energy tensor which contains the higher-order curvature terms has been used instead of the ordinary  stress-energy tensor, so the violation of NEC is due to the effective stress-energy tensor, not ordinary matter.  As an example in the brane-world scenario, four-dimensional brane is embedded in five-dimensional bulk  and the Einstein field equations are modified. The modification stems from higher-dimensional effect. In this scenario, ordinary matter  satisfies NEC and violation of NEC is due to terms coming from the bulk effects \cite{brane}. Wormholes in higher-dimensional spacetime have been investigated in the literature \cite{ndimen}. Some authors have studied the wormhole in the framework of  Brane-Dicke\cite{Dicke} and Lovelock\cite{Lovelock} which is considered as the most theory of gravitation in $n$ dimensions. In \cite{mehdi} authors have presented  some Einstein-Gauss-Bonnet traversable wormholes which satisfy  energy conditions. Wormhole Solutions in mimetic gravity \cite{mimet} and  Rastall gravity \cite{Rastall} have been investigated.
 Most of the modified theories of gravity have been used to resolve the problem of exotic matter in wormhole theories. Since none of the  modified gravity theories is  experimentally accepted as the superior theory. The efforts to find wormhole solutions in GR theory are more plausible.

Recent astrophysical observations   proposed a flat universe with an accelerated expansion \cite{1}. A barotropic  fluid with  an EoS, $p=\omega \rho $, with positive energy density  is a good candidate to explain the evolution of the cosmos. EoS parameter, $\omega$,  performs an important role to describe possible situations. The regime with $-1<\omega\leq 0$ is called dark energy and $\omega\leq -1$ is denoted as the phantom regime. Fluid with $\omega\leq -\frac{1}{3}$, can cause accelerated expansion of the Universe.  Since phantom fluid violates NEC, it can be considered as a suitable source to sustain wormholes. Several authors investigated wormhole with phantom energy \cite{phantom1,phantom2,phantom3,foad1}. Although
many of these solutions  \cite{phantom1,phantom2}, are not  asymptotically flat  in \cite{phantom3,foad1}, some asymptotically flat solutions have been presented. Jamill et al. have studied wormhole supported polytropic phantom energy which is a generalization of phantom energy and in some cases Chaplygin-gas models \cite{mobasher}. However  accelerated expansion of the Universe added plenty more doubts in the validation of energy conditions, violation of energy conditions is not completely acceptable. Therefore, minimizing  violation of energy conditions is yet very important in finding wormhole exact solutions.

Cosmological model  with variable EoS parameter has been studied in the literature \cite{variable1}.
This method can be used to sustain wormhole solutions with minimum violation of energy conditions. In \cite{rah}, wormholes supported by phantom energy with  variable EoS have been discussed. Since in  \cite{rah} EoS is considered in the phantom regime, violation of NEC is inevitable. Cattaldo and Orellana \cite{cat} assumed a shape function with a quadratic dependence on the radial coordinate. They have studied solutions with a vanishing redshift function. Their solutions consist of two part; a wormhole part and a dark energy part. In the wormhole part, the EoS has been considered variable but always  in the phantom regime. They have found some exact solutions which are not asymptotically flat. They have used cut and paste method to solve the problem of asymptotically flatness. In the cut and paste method, the interior wormhole solutions will be matched with an exterior Schwarzschild metric. It is a usual   method to resolve the problem of asymptotically flatness but seems to be nonphysically. The wormhole part of their solution violates NEC while the dark energy part does not. In another method \cite{lopez}, to minimize the violation of energy conditions, authors  assumed an EoS in which,  the sum of the energy density and radial pressure is proportional to a constant with a value smaller than that of the inverse area characterising the system. This approach is resulted to a class of solutions which are not asymptotically flat. Cut and paste method has been used to achieve asymptotically flat solutions. Garattini and Lobo have constructed some traversable wormholes supported by phantom energy, with an
$r$-dependent equation of state parameter. They have considered the possibility that these
phantom wormholes be sustained by their own quantum fluctuations \cite{Remo}.
In this work, we present some new wormhole solutions which are asymptotically flat and need small amount of exotic matter. We assume a variable EoS which tends to a constant parameter at large radial coordinate.  We present our solutions by considering an especial shape function.

 The organization of the paper is as follows: In the next section, the general equations and conditions of wormhole theory are presented. In
section \ref{sec3}, a specific shape function is presented and the possibility of  NEC violation is studied. Wormholes with a vanishing redshift function are investigated in section \ref{sec4}. We obtain exact wormhole solutions with minimum need to exotic matter for non vanishing redshift function in section \ref{sec5}, which are asymptotically flat. Concluding remarks are presented in the last section.


\section{Basic structure of wormhole theory}

The line element of a static and spherically symmetric wormhole is given by  \citep{WH}
\begin{equation}\label{1}
ds^2=-e^{2\phi(r)}dt^2+\left[ 1-\frac{b(r)}{r} \right]^{-1} dr^2+r^2\,d\Omega^2 \,,
\end{equation}
where $d\Omega^2= (d\theta^2+\sin^2\theta d\phi^2)$.  Here $b(r)$ is called the shape or form function, as it is related to the shape of the wormhole. The wormhole connects two different world or two distant part of the same universe at the  throat which is located at a minimum radial coordinate, $r_0$, with $b(r_0)=r_0$. There are some conditions on the shape function $b(r)$ :
\begin{equation}\label{f1}
\frac{(b-b' r)}{2b^2} > 0
\end{equation}
which is famous as the flaring-out condition, it reduces to $b'(r_0)<1$ at the wormhole throat (note that the prime denotes the derivative $\frac{d}{dr}$ and dot means the derivative $\frac{d}{dt}$). The condition $(1-b/r)>0$ is also imposed, so that $b(r)<r$, for $r>r_0$.
The function $\phi(r)$ is famous as the redshift function, which should be  finite everywhere to avoid the existence of horizon in the spacetime.
Asymptotically flat condition for these two functions leads to
\begin{equation}\label{f2}
\lim_{r\rightarrow \infty}\frac{b(r)}{r}=0 , \quad \lim_{r\rightarrow \infty}\phi(r)=0.
\end{equation}
In this work, we are interested in obtaining asymptotically flat geometries.
Due to the specific structure of  wormhole, the stress-energy tensor is not generally isotropic. So, we should consider the inhomogeneous property of the wormhole spacetime. In Ref. \cite{phantom1} the extension of phantom energy to inhomogeneous and anisotropic spherically symmetric
spacetimes has been studied.
We consider an anisotropic fluid in the form $T^\mu_\nu={\rm diag}(-\rho, p,p_t,p_t )$, where $\rho(r)$ is the energy density, $p(r)$ is the radial pressure, and $p_t(r)$ is the lateral pressure. Now, using the  Einstein field equations, we obtain the following distribution of matter (with $G=c=1$)
\begin{eqnarray}\label{2}
b'&=&8\pi r^2\rho, \\
\label{6}
\phi'&=&\frac{8\pi p r^3+b}{r(r-b)},\\
p_t&=&p+\frac{r}{2}\left[ p' + (\rho + p)\phi'  \right]\,.
\label{pt}
\end{eqnarray}
The conservation of the stress-energy tensor, $T^{\mu\nu}{}_{;\mu}=0$, also reproduces the equation (\ref{pt}).
Although the fluid for this line element is not isotropic, we can use the radial pressure in the  EoS, $p=f(\rho)$, which was first presented in the study of phantom wormhole solutions \cite{phantom1}. In this paper, we consider a linear like EoS as follows:
\begin{equation}\label{f3}
p=\omega_{eff}(r)\rho(r)=(\omega_\infty+g(r))\rho(r),
\end{equation}
here $\omega_{eff}(r)$ is the effective state parameter and  $\omega_\infty$, denotes the constant state parameter at the large radial coordinate.
In order to have an asymptotically linear EoS, the condition
\begin{equation}\label{f4}
\lim_{r\rightarrow \infty}g(r)=0,
\end{equation}
is also imposed.
Asymptotically flatness of  spacetime is an important property which should be taken into account in the study of  wormhole exact solutions. Many authors confined the wormhole solutions to some interior spacetime. Then, using junction conditions method, match these interior wormhole solutions to
the exterior Schwarzschild solution. We will study the wormhole solutions which are intrinsically asymptotically  flat, so it is not necessary to surgery the wormhole to an exterior Schwarzschild solution. In asymptotically flat spacetime at large radial coordinate, $\rho(r)$ and $p(r)$ should tend to zero. So by considering a well-defined function for EoS, we have
 \begin{equation}\label{9a}
 \lim_{\rho\rightarrow 0}p(\rho)=0.
\end{equation}
We can define a mass function
\begin{equation}\label{9b}
m(r)\equiv \int_{r_0}^r 4\pi r^2\rho dr.
\end{equation}
By taking into account Eq.(\ref{2}),
\begin{equation}\label{f10}
m(r)=\frac{b(r)-r_0}{2}.
\end{equation}
In\cite{visser}, a volume integral
\begin{eqnarray}\label{f11}
I_V\equiv \int (\rho+p_r)dV&=&\left[ (r-b)\ln \left(\frac{e^{2\phi(r)}}{1-b/r}\right)\right]^\infty_{r_0}
   \nonumber  \\
&& \hspace{-1.7cm} - \int_{r_0}^\infty
(1-b'(r))\left[ \ln \left(
\frac{e^{2\phi(r)}}{1-\frac{b}{r}}\right)\right] dr.
\end{eqnarray}
has been introduced  instead of averaged null energy condition to measure the amount of NEC violation. Zaslaveskii\cite{zas} has used the same integral but he assumed that the wormhole spacetime can be divided into two regions while the exotic matter exists in the inner region, $r_0<r\leq a$, the outer region, $r\geq a$, is filled by the normal matter. In this method, the  total amount of exotic matter measured by
\begin{equation}\label{f12}
I=8\pi \int_{r_0}^a (\rho+p)r^{2}dr .
\end{equation}
   In the next sections, we will use this integral to quantify the amount of NEC violation.
Considering an effective state parameter help us to investigate a large class of solutions which can be reduced to the famous solutions in the suitable limits. For example, solutions with $g(r)=0$, are related to wormholes with phantom EoS which have been studied in the literature \cite{phantom1,foad1}.

From now, we will study the possibilities to find new solutions. We have several functions, namely, $\phi(r)$, $b(r)$, $\rho(r)$, $p(r)$, $p_t(r)$
and EoS parameter  $\omega_{eff}$. On the other hand, we have three field equations, Eqs.(\ref{6})-(\ref{pt}) and Eq.(\ref{f3}) as EoS . To close the system, we should equal the  number of  equations with the number of unknown functions. We can use several strategies to solve the problem. For example, one can consider a special function for $\rho(r)$ or considering an extra condition on the lateral pressure \cite{phantom1}.
Using an arbitrary shape function or redshift function  is another method to close the system. All of these strategies lead to exact wormhole solutions and could be used in a reverse manner to accomplish each other. For instance, one can consider a known shape function, $b(r)$, but from Eq.(\ref{2}) it is clear that there is not any difference between choosing a known energy-density, $\rho(r)$, instead of $b(r)$. In the following, we will use a special form function to  find exact solutions.

\section{Wormholes with special shape function }\label{sec3}

A large number of wormhole solutions, which have been studied in the literature, deal with a shape function in the following form
   \begin{equation}\label{13}
 b(r)=A\,r^{\alpha}+h(r)
 \end{equation}
where, $A\,r^{\alpha}$, is the term with the highest-order of $r$ in the shape function and $h(r)$, is an arbitrary function with lower-order terms.
In general, the shape function can be classified into four  categories:
 the first category  is  the famous solution $b(r)=r_{0}$ and would not be considered here. The three other categories are related to positive, negative or mixed energy density.First, we will consider positive energy density to have more consistency with recent observations. For a strictly increasing shape function, $\rho(r)$ would be always positive. Asymptotically flat condition implies  $\alpha \leq 1 $ . The necessary conditions on $b(r)$ in Eq.(\ref{13}) to be a strictly increasing function, becomes
  \begin{equation}\label{13a}
A \geq 0 ,\qquad 1 \geq \alpha \geq 0.
\end{equation}
Eq.(\ref{9b}) implies that these restrictions on $b(r)$ will produce the  solutions with unbounded mass function.

  Let us study the possibility of violation of NEC. The $\omega_{eff}$ is a function which starts at the throat $(\omega_{0}<-1)$ and tends to $\omega_{\infty}$. To satisfying NEC, it must exceed $\omega_{eff}=-1$.
    Generally, wormhole with positive energy density, can satisfy NEC in some regions of space or at large distance from wormhole. In other words, three situations are possible for $\omega_{eff}(r)$ which are plotted in Fig.1.
  In case (a), $\omega_{eff}(r)$ never exceeds the line $\omega_{eff}=-1$. So the NEC is violated in the whole of space while in the case (b), $\omega_{eff}(r)$ exceeds $\omega_{eff}=-1$, at some $r=r_{1}$ and again at $r=r_3$. Therefore, the NEC is not violated only in the interval $r_{1}\leq r \leq r_{3}$ .
   In  case (c), NEC is not satisfied only in the interval $ r\leq r_{2} $ .
  In the next sections, we will try to find solutions which are compatible with case (c). This class of solutions seems to be more physical because the EoS is different from linear form only in the vicinity of the wormhole throat and tends to a linear EoS, in the recent part of the space. The motivation behind it, to explore such solutions, is that the EoS of Cosmos is a global equation and in local view the EoS may be different. Therefore, one can choose a variable EoS parameter instead of a constant EoS parameter. This provides possibilities to find solutions which satisfy NEC in some regions of spacetime.
   \begin{figure} [h t] \label{fig1}
\centering
  \includegraphics[width=3.2in]{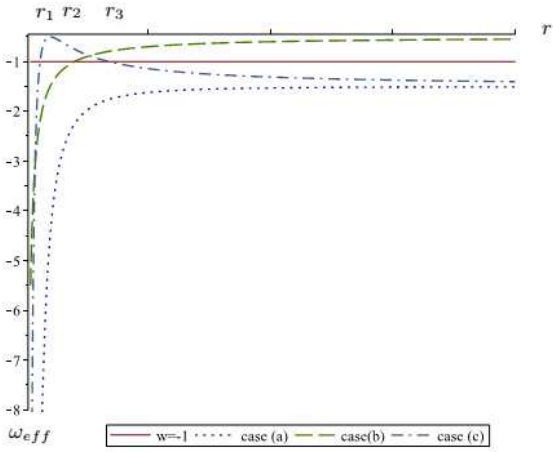}
\caption{Three possible cases for $\omega_{eff}$. In case(a) (dotted line), $\omega_{eff}$ is always smaller than the line $\omega_{eff}=-1$ (solid line) so  NEC is violated throughout the entire range of $r$. In case(b) (dotted-dashed line),   $\omega_{eff}$ is bigger than the line $\omega_{eff}=-1$ (solid line)in the range $r_{1}<r<r_{3}$ so  NEC is not violated throughout this range. In the case(c)(dashed line), $\omega_{eff}$ exceeds  the line $\omega_{eff}=-1$ (solid line) at $r=r_{2}$ and tends to a constant so  NEC is not violated throughout the range $r_{2}<r$.
 See the text for details.}
\end{figure}
  
\section{Wormholes with a vanishing redshift function}\label{sec4}

Vanishing redshift function ($\phi=0$) implies that there is not any  tidal force.  In this section, we analyse the wormhole with a vanishing redshift function. For this category of solutions from Eqs.(\ref{2})- (\ref{6}) one can find that
 \begin{equation}\label{10}
 p(r)=(\omega_{\infty}+g(r))\rho(r)=-\frac{b(r)}{8\pi r^3}
\end{equation}
and
 \begin{equation}\label{11}
\omega_{eff}(r)=-\frac{b(r)}{rb'(r)},
\end{equation}
on the throat
 \begin{equation}\label{12}
\omega_{eff}(r=r_{0})=\omega_{0}=-\frac{1}{b'
(r=r_{0})}.
\end{equation}
To check NEC, first we have
\begin{equation}\label{12f}
(1+\omega_{eff})\rho>0,
\end{equation}
also for vanishing redshift function from Eq.(\ref{pt}), we obtain
\begin{equation}\label{12d}
p_t=-\frac{p_r+\rho}{2}.
\end{equation}
Hence, the expression $\rho+p_t>0$ leads to
\begin{equation}\label{12e}
(1-\omega_{eff})\frac{\rho}{2}>0.
\end{equation}
By taking into account Eqs.(\ref{12f}),(\ref{12e}), we conclude that for $\rho>0$ these requirements, in turn, lead to the following restriction on $\omega_{eff}$ in order to satisfy NEC,
\begin{equation}\label{15a}
-1<\omega_{eff}<1.
\end{equation}
By considering Eq. (\ref{11})
\begin{equation}\label{12c}
\omega_{\infty}=\lim_{r\rightarrow \infty} w_{eff}(r)=\lim_{r\rightarrow \infty}-\frac{b(r)}{rb'(r)}=-\frac{1}{\alpha}.
\end{equation}
For the special  shape function, Eq.(\ref{13}), since $\alpha<1$, it shows that possible range for $\omega_{\infty}$ is as follows
 \begin{equation}\label{14a}
\omega_{\infty}<-1 \quad or \quad \omega_{\infty}>0 .
\end{equation}
It is also of a particular interest to analyse the NEC at large distance from wormhole.
By taking into account Eq.(\ref{13a}), one can deduce that solutions which satisfy NEC at large radial coordinate are not possible for the special shape function, considered in  Eq.(\ref{13}), with a vanishing redshift function.
So solutions with the special shape function Eq. (\ref{13}) and $\phi=0$ can be assumed as an example for case (a) or (b) in Fig.1 but not for case (c).

\section{Wormhole with non vanishing redshift function}\label{sec5}

For non vanishing redshift function $\omega_{eff}$ is
\begin{equation}\label{18a}
 \omega_{eff}=-\frac{b}{rb'}+\frac{2\phi'(r-b)}{\,b'}.
 \end{equation}
 We are interested in traversable  wormhole with no horizon, so $\phi(r)$ should be finite everywhere. Eq.(\ref{18a}) indicates that $\phi(r)$ contributes in the $\omega_{\infty}$ if
  \begin{equation}
  \lim_{r\rightarrow \infty}\frac{\phi'\,(r-b)}{b'}=finite.
 \end{equation}
   Since $b(r)<r$, it reduces to
 \begin{equation}\label{19a}
 \lim_{r\rightarrow \infty}\frac{\phi'\,r}{b'}=D,
 \end{equation}
 where $D$, should be a finite constant. Then it is easy to show that $\phi$ must be in the form
  \begin{equation}\label{20a}
\phi(r)=A_{1}\,r^{\alpha-1}+s(r)
\end{equation}
while
\begin{equation}\label{20aa}
A_{1}=\frac{D\alpha A}{\alpha-1},
\end{equation}
 is a constant and  $s(r)$ could be a general function which its order is less than $r^{\alpha-1}$. From Eq. (\ref{18a}), it is easy to show that contribution of redshift  and shape functions in the $\omega_{\infty}$ is
\begin{equation}\label{21a}
\omega_{\infty}= \lim_{r\rightarrow \infty}\omega_{eff}=-\frac{1}{\alpha}+2D.
 \end{equation}
 It shows that there is a relation between $\omega_{\infty}$ and the largest term in the redshift and shape functions. Eq.(\ref{21a}) implies that if  $A_{1}=0$, then $\phi$ has not influence on the $\omega_{\infty}$. Therefore similar to vanishing redshift function, in this situation, it is impossible to find solutions in which NEC is satisfied at large distance from wormhole.
 Using Eq.(\ref{f3})  and some calculations one can shows
\begin{equation}\label{22a}
g(r)=\frac{\omega_{\infty}h'r+s'r^{2}-\phi'rb-h}{rb'}.
 \end{equation}
  We can use this equation to analyse the rate of convergence of $g(r)$ to zero. The term, $\frac{s'r^{2}}{rb'}$ in Eq.(\ref{22a}), implies that $s$ should be equal to zero or its order must be less than $r^{\alpha-2}$.

   Now, we have the essential mathematical tools to find solutions which do not satisfy $NEC$ in some region near the wormhole throat, but for  some $r_{2}<r$  this solutions satisfy NEC (Fig.1 case (c)).
We consider $b(r)$ as follow
\begin{equation}\label{23a}
b(r)= r_{0}(A (\frac{r}{r_{0}})^{\alpha}+1-A).
 \end{equation}
  From Eq.(\ref{20a}), we conclude that
 \begin{equation}\label{24a}
\phi(r)=A_{1} (\frac{r}{r_{0}})^{\alpha-1}+s(r).
 \end{equation}
 \begin{figure} \label{fig2}
\centering
  \includegraphics[width=3.2in]{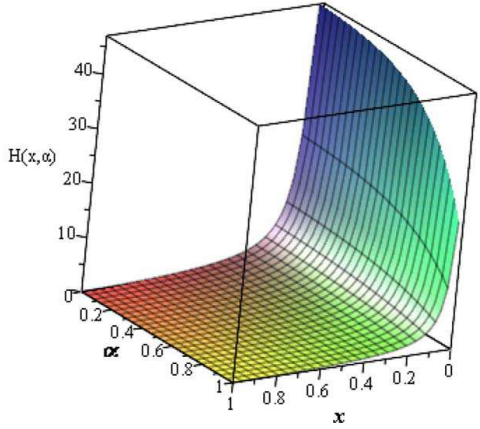}
\caption{The plot depicts the function $H(x,\alpha)$ for $A=\frac{1}{2}$ against $x$ and $\alpha$ . It is clear that $H(x,\alpha)$ is positive throughout the range  $0<x<1$ and $0<\alpha<1$, which implies $b(r)<r$ is satisfied everywhere. See the text for details.}
\end{figure}
For the sake of simplicity, we set $s(r)=0$.
 From now, we define the dimensionless parameter $x\equiv\frac{r_{0}}{r}$ which has the range $0<x\leq 1$. Note that $x=1$ corresponds to the throat and $x\longrightarrow 0$ corresponds to spatial infinity. The shape function and the redshift function take the form
\begin{equation}\label{23b}
B(x,\alpha)=\frac{b(r)}{r_{0}}=(A\,x^{-\alpha}+1-A)
 \end{equation}
 and
 \begin{equation}\label{24b}
\phi(x)=A_{1} x^{(1-\alpha)}.
 \end{equation}
To check the condition $b(r)<r$, one can define
\begin{equation}\label{24ab}
H(x,\alpha)= \frac{1}{x}-B(x,\alpha).
 \end{equation}
As an example, we have plotted $H(x,\alpha)$ for $A=\frac{1}{2}$ as a function of $x$ and $\alpha$  in Fig.2 which indicates that $H(x,\alpha)$ is positive through the entire range $0<x<1$ and $0<\alpha<1$. So the condition $b(r)<r$ is satisfied everywhere. Since $B(x=1)=1$ and $\lim_{x\rightarrow 0}B(x)=0 $,  one can conclude that this shape function has the essential conditions to construct the asymptotically flat wormhole solutions.

 According to \cite{foad1}, the gravitational redshift as measured by a distant observer, is given by
\begin{equation}
z=\frac{\delta \lambda}{\lambda}=1-\frac{\lambda (r\rightarrow \infty
)}{\lambda(r=r_0)}=\frac{1}{\exp( \phi (x=1)) }.
\end{equation}
One can use this to show that
\begin{equation}\label{25ba}
A_{1}=-\ln (1-z).
 \end{equation}
This equation implies the relation between  gravitational redshift, as detected by a distant observer, and coefficient $A_{1}$ which can be connected to the shape function (Eq.(\ref{20aa})). From Eqs.(\ref{2})-(\ref{pt}), one can deduce that the energy density
\begin{equation}\label{25a}
\rho(x)=\frac{A \, \alpha \,x^{3-\alpha}}{8 \pi r_{0}^{2}}
 \end{equation}
 is always positive. Radial and lateral pressure are
  \begin{eqnarray}\label{25c}
p(x)&=&-\frac{(a_{1}x^{3} +a_{2}x^{4-\alpha} +a_{3}x^{4-2\alpha} +a_{4}x^{3-\alpha})}{8\pi r_{0}^{2}}\\
p_{t}(x)&=&-\frac{1}{8\pi r_{0}^{2}}(c_{1}x^{3} +c_{2}x^{4-\alpha} +c_{3}x^{3-\alpha}
   \nonumber  \\
&&\hspace{1cm}+c_{4}x^{5-2\alpha}+c_{5}x^{4-2\alpha} +c_{6}x^{5-3\alpha}).
\end{eqnarray}
where
 \begin{eqnarray}\label{25cc}
a_{1}&=&(A-1),\quad a_{2}=2\,A_{1}\,(1-\alpha)(1-A) \nonumber \\
a_{3}&=&2\,A_{1}\,(1-\alpha)A,\quad a_{4}=2\,A_{1}\,(1-\alpha)-A
 \end{eqnarray}
 and
 \begin{eqnarray}\label{25ccc}
c_{1}&=&\frac{(A-1)}{2},\quad c_{2}=\frac{A_{1}(1-A)\,(3-5\alpha+2\,\alpha^{2})}{2} \nonumber \\
c_{3}&=&\frac{A(\alpha-1)-2\,A_{1}(\alpha-1)^{2}}{2},\quad \nonumber \\
c_{4}&=&2\,A_{1}^{2}\,(1-A)(\alpha-1)^{2},\nonumber \\
c_{5}&=&\frac{(3A-2A_{1})A_{1}\,(\alpha-1)^{2}}{2},\nonumber \\
c_{6}&=&A^{2}_{1}\,(\alpha-1)^{2}.\nonumber \\
\end{eqnarray}
\begin{figure} \label{fig3}
\centering
  \includegraphics[width=3.2in]{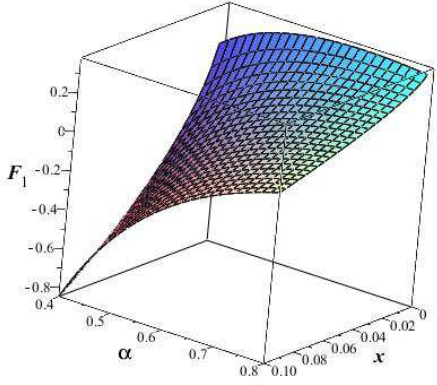}
\caption{The plot depicts the function $F_{1}(x,\alpha)$, for $A=\frac{1}{2}$ . It is transparent that $F_{1}(x,\alpha)$ is positive throughout
some range of $x$ and $\alpha$, which implies NEC is satisfied through this range. See the text for details.}
\end{figure}
\begin{figure} \label{fig4}
\centering
  \includegraphics[width=3.2 in]{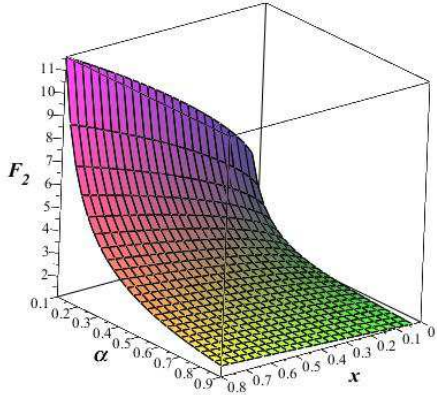}
\caption{The plot depicts the function $F_{2}(x,\alpha)$, for $A=\frac{1}{2}$ . It is transparent that $F_{2}(x,\alpha)$ is positive throughout the entire range of $x$ and $\alpha$, which implies NEC is satisfied. See the text for details.}
\end{figure}
  \begin{figure} \label{fig5}
\centering
  \includegraphics[width=3 in]{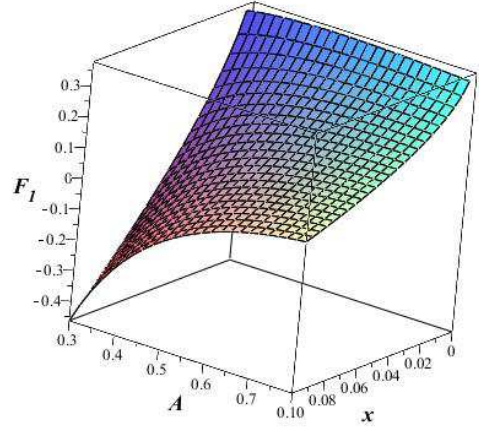}
\caption{The plot depicts the function $F_{1}(x,A)$, for $\alpha=\frac{2}{3}$ . It is clear that $F_{1}(x,A)$ is positive throughout some range of $x$ and $A$, which implies NEC is satisfied through this range. See the text for details.}
\end{figure}
\begin{figure} \label{fig6}
\centering
  \includegraphics[width=3 in]{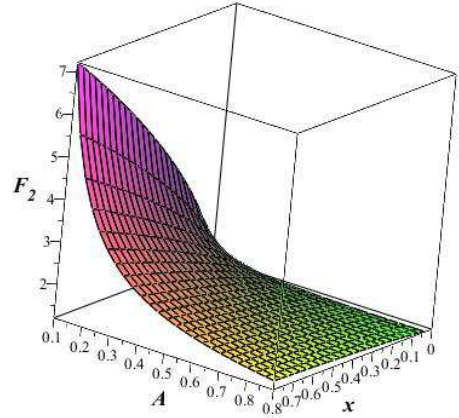}
\caption{The plot depicts the function $F_{2}(x,A)$, for $\alpha=\frac{2}{3}$ . It is clear that $F_{2}(x,A)$ is positive throughout entire range of $x$ and $A$ which implies NEC is satisfied. See the text for details.}
\end{figure}
Eq.(\ref{25c}) shows that
\begin{equation}\label{R0}
\alpha=1+\frac{8\pi r_0^{2}p_0-1}{4A_1}
 \end{equation}
where $p_0$ is the radial pressure at the throat of the wormhole. Using Eqs.(\ref{R9}) and (\ref{25ba})
 presents $\alpha$  in  terms of the known physical quantities. Eq.(\ref{25a}) helps us to define $A$ in the following form,
\begin{equation}\label{R01}
A=\frac{8\pi r_0^{2}\rho_0}{\alpha}
 \end{equation}
where $\rho_0$ is the energy density at the throat of the wormhole.
\begin{figure}\label{fig7}
\includegraphics[width=3 in]{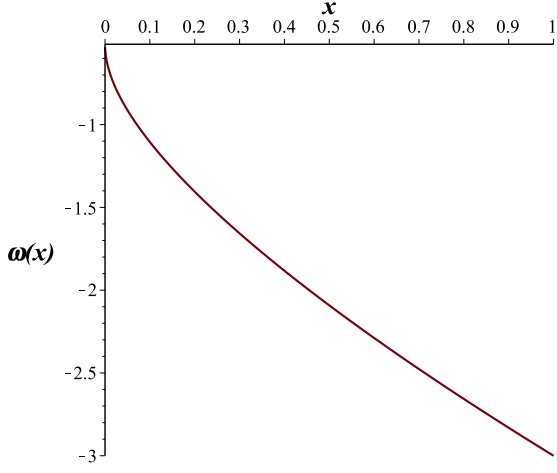}
\caption{Plot of $\omega(x)$ for $\alpha=2/3$ and $A=1/2$. It shows that  $\omega(x)$ exceeds $\omega=-1$  throughout the region $0\leq x \leq 0.071$ . It corresponds to case (c) in Fig.1. See the text for details.}
\end{figure}
To check violation of energy condition, we introduce functions
\begin{equation}\label{266a}
F_{1}(x,A,\alpha)=1+\frac{p(x)}{\rho (x)}
 \end{equation}
and
\begin{equation}\label{266aa}
F_{2}(x,A,\alpha)=1+\frac{p_{t}(x)}{\rho (x)}.
 \end{equation}
 Since $\rho(x)$ is always positive, if the sign of these two functions are positive in some regions, NEC is satisfied in those regions.
In general, to check the violation of NEC, we should plot  $F_{1}$ and $F_{2}$ as a function of $x$ and $\alpha$ for a specific choice of the parameter $A$. Also we can plot $F_{1}$ and $F_{2}$  against $x$ and $A$ for a specific choice of the parameter $\alpha$ . We have plotted $F_{1}$  and $F_{2}$ as a function of $x$ and $\alpha$ for $A=\frac{1}{2}$ in Fig.3 and Fig.4. We have plotted $F_{1}$  and $F_{2}$ as a function of $x$ and $A$ for $\alpha=\frac{2}{3}$ in Fig.5 and Fig.6. It is clear that in some regions for $x$ and $A$ the sign of $F_{1}$ and $F_{2}$ are simultaneously positive. From Fig.4 and Fig.6 it seems that $F_{2}$ is always positive in the whole regions for selected parameter. As it was mentioned, when the sign of $F_{1}$ and $F_{2}$ are simultaneously  positive, the NEC is satisfied. So, one can conclude that the sign of $F_{1}$ describes, the violation on NEC. It is interesting to note that  Fig.3 shows that for a constant $A$ as $\alpha$ increases( or $x$ decreases), $F_{1}$ increases  in which physically means, the phantom region will approach to the vicinity of the wormhole throat as $\alpha$ increases. Fig 5 presents the same behavior for $A$, this implies that for a constant $\alpha$ as $A$ increases, the phantom regime will approach  more to the vicinity of the wormhole throat.
Let us seek a special example for wormhole solutions violating NEC only in the vicinity of the wormhole throat in details. Fig.3-6 help us to consider $A=-A_1=\frac{1}{2}$ and $\alpha=\frac{2}{3}$ , which lead to the line element
\begin{equation} \label{28a}
ds^2=- e^{-(\frac{r_{0}}{r})^{\frac{1}{3}}}dt^2
+\frac{dr^2}{1-\frac{r_0}{2r}\left((\frac{r}{r_0})^{\frac{2}{3}}+1\right)}
+r^2\,d\Omega^2.
\end{equation}
The stress energy tensor components become
\begin{eqnarray}\label{29}
\rho(x)&=&\frac{\left(x \right)^{\frac{7}{3}}}{24\pi r_0^2} ,\\
p(x)&=&-\frac{(x^{\frac{7}{3}}+x^{\frac{10}{3}}+x^{\frac{8}{3}}+3\,x^{3})}{48\pi r_0^2}  ,\\
p_{t}(x)&=&-\frac{(17\,x^{3}+2\,x^{\frac{7}{3}}+5\,x^{\frac{10}{3}}+5\,x^{\frac{8}{3}}-x^{\frac{11}{3}})}{576\pi r_0^2}  .
\end{eqnarray}
 One can see that $\omega(x)$ take the form
\begin{equation}\label{28b}
\omega(x)=-(\frac{1}{2}+\frac{x+x^{\frac{2}{3}}+x^{\frac{1}{3}}}{2}).
 \end{equation}
We have plotted $\omega(x)$  as a function of $x$ in Fig.7 . This graph shows that for $ x\leq x_{1}$, where $x_{1}= (\sqrt{2}-1)^{3}\simeq 0.071$, EoS parameter $\omega$  is more than $-1$. This  corresponds to case (c) in Fig.1. It is obvious  that $\omega_{\infty}=-\frac{1}{2}$.
We have ploted  $F_{1}$ and $F_{2}$ as a function of $x$ in Fig.8. Since $\rho(r)$ is positive this figure demonstrates that in the interval $0\leq x \leq x_{1}$  $NEC$ is not violated.
\begin{figure}\label{fig8}
\includegraphics[width=3.2in]{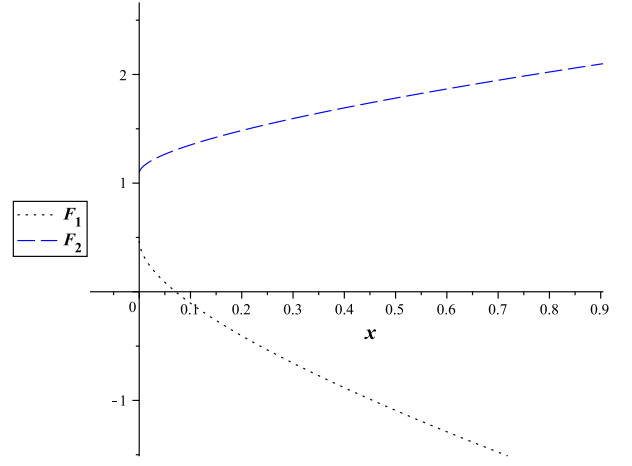}
\caption{The plot depicts the function $F_{1}(x)$ and $F_{2}(x)$  for $\alpha=2/3$. The function $F_{2}(x)$ (dashed line) is positive in the entire spacetime and it is transparent that $F_{1}(x)$ (doted line) is negative throughout the region $x>x_{0}$ when $x_{0}\simeq 0.071$, so the exotic matter is confined in this region. See the text for details.}
\end{figure}

One can evaluate volume integral quantifier, Eq.(\ref{f11}). For the metric (\ref{28a}), $I_V\longrightarrow +\infty$, which means the total amount of exotic matter is equal to zero. On the other hand, one can use Eq.(\ref{f12}) instead of Eq.(\ref{f11}).
This seems to be a better integral to provide information about the ``total amount'' of energy-condition-violating matter in the spacetime. Since $a=\frac{r_{0}}{x_{1}}$, the total amount of exotic matter threading the present wormhole solution is given by
\begin{equation}\label{f13}
I=8\pi \int_{r_0}^a (\rho+p)r^{2}dr =-1.114\,r_{0}.
\end{equation}
 To summarize, the line element Eq.(\ref{28a}) is an exact wormhole solution with a variable EoS parameter which its minimum is at the throat ($\omega_{0}=-3$) and exceeds $\omega=-1$ at $r_{2}=\frac{r_{0}}{x_{1}}=14.08\,r_{0}$  which means, the wormhole geometry is in the phantom era in the vicinity of the throat and is connected to a distribution of dark energy at $r=r_{2}$.
\subsection{ bounded mass wormhole }
Let us study solutions with a bounded mass function. We propose a shape function as follows
\begin{equation}\label{R1}
b(r)= r_{0}(L (\frac{r_{0}}{r})^{\alpha_{1}}+1-L).
 \end{equation}
In order to have a finite mass, $\alpha_{1}\geq 0$ is imposed. Eq.(\ref{f10}) indicates that
 \begin{equation}\label{R2}
m_{\infty}=\lim_{r\rightarrow \infty}m(r)=-\frac{L}{2}r_0
 \end{equation}
 where $m_{\infty}$ is a finite mass as seen by a distant observer.
 The relevant energy density
  \begin{equation}\label{R3}
\rho(r)=-\frac{\alpha_{1}L}{8\pi r^2_0} (\frac{r_{0}}{r})^{\alpha_{1}+1}.
 \end{equation}
 implies that to have a positive $\rho$, $L$ must be negative. Now, we will discuss  $\omega_{\infty}$. It is obvious that $\frac{b}{r\,b'}=-\frac{1}{\alpha}+\frac{1-L}{r\,b'}$. Since, $\lim_{r\rightarrow \infty}\frac{1-L}{r\,b'}\longrightarrow \infty$
, we cannot complete the solution following the previous procedure. Using Eqs.(\ref{18a}) and (\ref{R1}),
 \begin{equation}\label{R4}
\phi(r)=\frac{1}{2}\int \frac{g(r)L\,\alpha_1 +\omega_\infty L\,\alpha_1+(L-1)r^{\alpha_1} -L }{r((L-1)r^{\alpha_1}+r^{1+\alpha_1}-L)}dr
 \end{equation}
is achieved. Solving this integral for a general $\alpha_1$ and $g(r)$ is difficult. Thus, we put $\alpha_1 =-1$ and $g(r)=\frac{-q\,r^2_0}{r^2}$ (with $q>0$) which results
 \begin{eqnarray}\label{R5}
 \phi(r)=(\frac{qL-\omega_\infty L+1}{2(L+1)})\ln(\frac{r}{r_0}-1)\nonumber \\
+\frac{L-\omega_\infty +\frac{q}{L^2}}{2(\frac{r}{r_0}+L)}\ln(\frac{r}{r_0}+L)+\frac{qr^2_0}{4r^2}+\frac{qr_0}{2r}(1-\frac{1}{L}).
 \end{eqnarray}
To avoid from having horizon in throat ($ \lim_{r\longrightarrow r_0}\phi(r)\longrightarrow \infty$), we should set
 \begin{equation}\label{R5a}
q=\omega_\infty -\frac{1}{L} \,, \qquad \qquad L > -1.
 \end{equation}
which leads to the line element
\begin{eqnarray} \label{R7}
ds^2&=&-(1+\frac{L}{r})^{d_1} e^{-(\frac{d_{2}r_0}{r}+{\frac{d_3 r^2_{0}}{r^2}})}dt^2 \nonumber \\
&+&\frac{dr^2}{1-\frac{r_0}{r}\left((\frac{L\,r_0}{r})+1-L\right)}+r^2\,d\Omega^2,
\end{eqnarray}
where
\begin{eqnarray} \label{R7a}
d_1&=&\frac{L^3- (1+\omega_\infty)L^2+(1+\omega_\infty)L-1}{L^3},  \nonumber \\
d_2&=& \frac{\omega_\infty L^2- (1+\omega_\infty)L+1}{L^2} ,\nonumber \\
d_3&=& \frac{\omega_\infty L -1}{2L}.
\end{eqnarray}
The stress energy tensor components for $q=\frac{3}{2}, L=-\frac{1}{2}$ (an special case)  become
\begin{eqnarray}\label{R8}
\rho(x)&=&\frac{1}{16\pi r_0^2}(x^2) ,\\
p(x)&=&- \frac{1}{32\pi r_0^2}(x^2+3x^4) ,\\
p_{t}(x)&=&\frac{3}{256\pi r_0^2}(\frac{2x^3+17x^4-13x^5-3x^7}{2-x})  .
\end{eqnarray}
 Let us study the violation of NEC. It is obvious that $p(r)+\rho(r)$ is negative from the throat up to $r=r_2=a$ (see Fig.3 case(c)) where $a$ is the root of $\omega(a)=-1$. From Eq.(\ref{R5a}), it is clear that
\begin{equation}\label{R9}
a=\frac{q}{1+\omega_\infty}=\frac{\omega_\infty -\frac{1}{L}}{1+\omega_\infty}  .
\end{equation}
Eq.(\ref{R5a}) indicates that $L$ has a minimum. If we consider $L=-1+\epsilon$ as  the lowest possible value for $L$ ($\epsilon$ is a  small value), then one can show
\begin{equation}\label{R10}
a=1+\frac{1}{2(1+\omega_\infty)}\epsilon +O(\epsilon^2)  .
\end{equation}
so
\begin{equation}\label{R11}
I=8\pi \int_{r_0}^a (\rho+p)r^2dr= \frac{-\epsilon^2}{2(1+\omega_\infty)}+O(\epsilon^3)) .
\end{equation}
To summarize, one can consider the line element (\ref{R7}) as a wormhole with mass, $ m = \frac{1-\epsilon}{2}r_0$, as detected by a distant observer. The stress energy due this wormhole violate NEC in the interval $r_0\leq r <r_{0} (1+\frac{1}{2(1+\omega_\infty)}\epsilon)$ with the total amount of NEC violation $I= \frac{-\epsilon^2}{2(1+\omega_\infty)}$. These results can be achieved for other forms of $g(r)$ and $b(r)$ by carefully fine-tuning the parameters to find $\phi(r)$.

 \subsection{wormhole with a mixed energy density }\label{subsec1}
 Considering the stress-energy, in the vicinity of wormhole throat, in the phantom regime is a possible candidate for exotic matter. Someone may prefer to consider another form of exotic matter in this region. So, one should work with a negative energy density instead of a positive one. As it was mentioned before, solutions with positive $\rho$ is more acceptable in the dark energy era. Matter with $\rho<0$ violates dominant and weak energy conditions. So, the sign of $\rho$ should be changed in the boundary between exotic era and dark energy era. The possible candidate is a mixed energy density.
To construct a wormhole with a mixed energy density, we start with the shape function of Eq.(\ref{13}).
From Eq.(\ref{2}), it can be verified that $b(r)$ should have a minimum in which $\rho'=b'=0$. The sign of $\rho$ changes in this point. So we should consider a shape function with a minimum, i.e.
\begin{equation}\label{R12}
b(r)=(A_3(r/r_0)^{\alpha_2}+2A_3(r/r_0)^{\alpha_3}+(1-3A_3))r_0 .
\end{equation}
 It is easy to show that the non vanishing root of $\rho(r)=0$ is
 \begin{equation}\label{R13}
r^*=r_0(\frac{2\alpha_3}{\alpha_2})^{\frac{1}{\alpha_2 -\alpha_3}} .
\end{equation}
\begin{figure}\label{fig9}
\includegraphics[width=2.7 in,height=2.4 in]{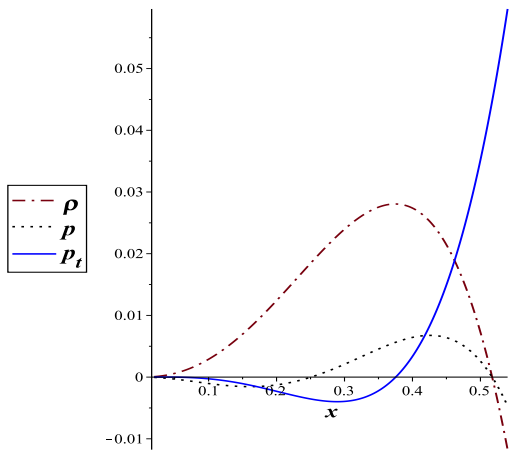}
\caption{The plot depicts the function $\rho(x)$ (dot-dashed line), $p(x)$ (dot line) and $p_t(x)$ (solid line)  for $A=-\alpha_{3}=1, \alpha_{2}=2/3$ and $A_{1}=-\frac{3^{1/5}\times(5\times3^{2/5}-6)}{2\times(3^{3/5}-5\times3^{2/5}+6)}\simeq-0.8136$ (note that vertical axe is scaled in $\frac{1}{8\pi\,r^2_0}$). The function $\rho(x)$ is positive in the range $x>x^*$ where $x^*\simeq 0.517$. See the text for details.}
\end{figure}
Since $\alpha_2>0$, the condition $\alpha_3<0$ is imposed to have a positive real root. Now we should try to find a suitable $\phi$. It was verified that to have a contribution in the $\omega_\infty$, $\phi$ must be in the form of Eq.(\ref{20a}). For the sake of simplicity, we put $s(r)=0$ in $\phi$. The $\omega_{eff}$ should be finite everywhere. Since $b'(r^*)=0 $, we must set
 \begin{equation}\label{R14}
 \omega_{eff}(r^*)=0,
 \end{equation}
 to find $A_1$. We do not state the form of $A_1$ here for abbreviation. We continue by putting $A=-\alpha_{3}=1, \alpha_{2}=2/3$  which leads to $A_{1}=-\frac{3^{1/5}\times(5\times3^{2/5}-6)}{2\times(3^{3/5}-5\times3^{2/5}+6)}\simeq-0.8136$. It is worth mentioning that Eq.(\ref{R14})  restricted the value of $A_1$. The energy-tensor, need to construct this geometry, is
 \begin{eqnarray}\label{R15}
\rho(x)&=&\frac{x^{\frac{7}{3}}-3\,x^4}{12\pi r_{0}^{2}} ,\\
p(x)&=&\frac{1}{12\pi r_{0}^{2}}((\frac{3}{2}-A_1)x^{\frac{7}{3}}\nonumber \\
&+&3x^{4}-3x^3-A_1x^{\frac{13}{3}}+2x^{\frac{10}{3}}) .
\end{eqnarray}
We do not state $p_t(x)$ here for abbreviation. The $\rho(x)$, $p(x)$ and $p_t(x)$ have been plotted as a function of $x$ in Fig.9
To check the violation of NEC, we have plotted $F_1(x)=8\pi r^2_0(\rho(x)+p(x))$ and  $F_2(x)=8\pi r^2_0(\rho(x)+p_t(x))$ as a function of $x$ in Fig.9 which implies that NEC is violated only in the region $x>x^*$ with $x^*\simeq 0.517$. To summarize, we have presented an exact wormhole solution with a variable EoS parameter which its energy density  is negative in the range $r_0\leq r< r^*$ and positive through $r^*<r$. The wormhole  is in the exotic era (not essentially phantom)  in the vicinity of the throat and is connected to a distribution of dark energy at $r=r_1$.
\section{Concluding remarks}
Cosmos with a linear  EoS is the most acceptable theory in recent cosmology. In this paper, by considering a general EoS which has reduced to an  asymptotically linear EoS and also assuming asymptotically flat conditions, some classes of possible wormhole solutions have been investigated. We have used a special shape function, $ b(r)= r_{0}((A \frac{r}{r_{0}})^{\alpha}+1-A)$, which can present a large class of solutions.
\begin{figure}\label{fig10}
\includegraphics[width=2.7 in,height=2.4 in]{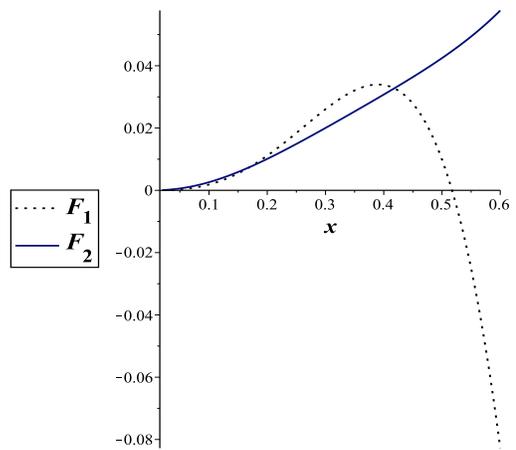}
\caption{The plot depicts the function $F_1(x)$ (dot line) and $F_2(x)$ (solid line)  for $A=-\alpha_{3}=1, \alpha_{2}=2/3$ and $A_{1}=-\frac{3^{1/5}\times(5\times3^{2/5}-6)}{2\times(3^{3/5}-5\times3^{2/5}+6)}\simeq-0.8136$ . The function $F_1(x)$ and $F_2(x)$ are positive in the range $x>x^*$ where $x^*\simeq 0.517$. So, NEC is satisfied in this region See the text for details.}
\end{figure}
 We have focused on solutions with positive energy
  density, which reduce the exoticity  of the fluid, and is more considered with nowadays cosmology. We have shown that some significant physical quantities; EoS parameter at large radial coordinate, $\omega_{\infty}$, gravitational redshift as measured by a distant observer, $z$, energy density at the throat, $\rho_{0}$, and peruser at the throat, $p_{0}$, are dependent to coefficients which have appeared in redshift  and shape functions.
Violation of NEC, as a fundamental ingredient of wormhole geometries, has been analysed. It has been shown that solution with vanishing redshift  function, which satisfies NEC at large radial coordinate for the selected shape function, does not exist. For non constant redshift function, we have studied solutions which are asymptotically flat and violate NEC in a small region in the vicinity of the wormhole throat. In this class of solutions, fluid near the wormhole throat is in the phantom regime and at some $r=r_{2}$, the phantom regime( or other exotic regime)  is connected to a dark energy regime.
 The boundary is  sensitive to change in parameters of the shape and redshift  functions . The total amount of exotic matter has been calculated for  solutions. This value is changeable because the region with exotic matter  can be controlled by fine-tuning the parameter.
We have presented a general formalism to find exact solutions, which more widespread than the previous formalism, and this is due to considering a variable EoS which can change from phantom era to dark energy era intrinsically. Variable EoS seems to be more physically since the linear EoS is a global equation and in local view, it is not necessary to  consider a  linear equation. One can relate this to the special geometry of wormhole near the throat. Variable EoS parameter, in contrast to constant one, has brought new life in studying wormhole physics.
 Although, in \cite{Remo} an asymptotically flat solution with variable EoS is presented. In the other works, the EoS is always in the phantom region \cite{rah} or solutions have a finite size and  a surgery is essential to glue them to an exterior solutions \cite{cat,lopez}. What is of particular interest is that in contrast to previous solutions \cite{rah,cat,lopez} our solutions are arbitrarily large. Meanwhile, there is  not any necessity to use cut and past method to solve the problem of asymptotically flatness.

As it was mentioned,  a flat space is connected to the original spacetime at a given hypersurface in cut and paste method. Generally, the Israel junction conditions should be applied in this surgery. Usually a surface stress-energy tensor is essential. A surface with no surface energy terms is famous as boundary surface while surface with stress-energy terms is called thin-shell\cite{lem}. The surface tangential pressure in thin-shell wormholes holds against collapsing or expansion of the boundary.  The first junction condition is the continuity of the metric components at surface($r=r_{s}$) , i.e. $g^{int}_{\mu \nu}(r_{s})=g^{ext}_{\mu \nu}(r_{s})$. One can consider  $g^{ext}_{rr}=(1-b^{ext}(r)/r$, where the function $b^{ext}(r)$ is different from the internal shape function ($b^{int}(r)$). When Eq.(\ref{f1}) is not satisfied or NEC is violated in the whole spacetime, the suitable choice for $b^{ext}(r)$ can resolve the problem. For example, $b^{ext}(r)=2m$  is a good candidate which may lead to an exterior Schwarzschild solution. Eq.(\ref{2}) explains that a geometry with an especial energy-tensor is connected to a geometry with a different energy-tensor in cut and paste method. In other words, two different solutions of Einstein filed equations are connected to construct an exact solution. This is the main difference between a spacetime constructed  by cut and paste method and an intrinsically asymptotically flat spacetime. For intrinsically asymptotically flat spacetimes, the form of stress-energy tensor is not changed in the whole of spacetime. The Smooth functions for $\rho(r)$ and $p(r)$ are more considerable. Thus we have preferred to work with the intrinsically asymptotically flat spactimes. In this article, we carefully have constructed a specific shape function by considering asymptotically flatness condition. The selected shape function has lead to a smooth energy density profile, possessing a maximum at the throat and vanishing at spatial infinity (case with $\rho>0$) or possessing a minimum at the throat and vanishing at spatial infinity(mixed energy density) . Variable EoS has supported us to control the violation of NEC.  In our method and the other variable EoS methods, using cut and paste, the value of NEC violation is controllable by fine-tuning the parameters related to the shape and redshift functions. Comparing results of section \ref{subsec1} with cut and past methods (see \cite{cat} or \cite{visser}) shows that, there is not any priority between our method and the other methods to quantify the violation of NEC.

  This method is flexible and can be used to find solutions which are asymptotically flat but always in phantom regime (See case (a) in Fig.1).
We have used a specific shape function to find exact solutions but one can use the formalism of this paper with other forms of shape function to find new solutions. Although there is no observational evidence result to existence of wormholes, astrophysical observations of  supernovae of type Ia and cosmic microwave background have opened a new window to studying wormholes theoretically. So theoretical researches, which suggest a minimum violation of NEC, are of great interest to help an advanced civilization to construct  wormholes or finding experimental evidence for wormholes.

\section*{Acknowledgements}
This work has been in part supported by a grant from the Research Council of Sirjan University of Technology.

\end{document}